\documentclass{ctr_summer}

\usepackage{ctrfont}
\usepackage{natbib}
\usepackage{undertilde}
\usepackage{color}

 \usepackage{graphicx}
 \usepackage{psfrag}
 \usepackage{epsfig}
\usepackage{amsmath,rotating}
\usepackage{dashrule}
\usepackage{undertilde}

\let\ig\includegraphics
\let\tw\textwidth
\DeclareUnicodeCharacter{2661}{FIX ME!!!!}

\usepackage{url}

\usepackage{color}

\usepackage{xcolor}
\newcommand\look[1]{\textbf{#1}}
\usepackage{tikz}
\newcommand{\tikzxmark}{%
\tikz[scale=0.23] {
    \draw[line width=0.7,line cap=round] (0,0) to [bend left=6] (1,1);
    \draw[line width=0.7,line cap=round] (0.2,0.95) to [bend right=3] (0.8,0.05);
}}
\newcommand{\checkmark}{%
\tikz[scale=0.23] {
    \draw[line width=0.7,line cap=round] (0.25,0) to [bend left=10] (1,1);
    \draw[line width=0.8,line cap=round] (0,0.35) to [bend right=1] (0.23,0);
}}

\title{Wall-modeled large-eddy simulation based on building-block flows}

\shorttitle{WMLES based on building-block flows}

\author{Y. Ling\footnote[1]{\label{Aff}Department of Aeronautics and Astronautics, Massachusetts Institute of Technology}, G. Arranz{\ref{Aff}}, E. Williams{\ref{Aff}}, K. Goc, K. Griffin \and A. Lozano-Durán{\ref{Aff}}  }

\shortauthor{Ling {\normalfont et~al.}}

\begin{document}
\pagenumbering{gobble}

\setcounter{page}{1}

\maketitle

A unified subgrid-scale (SGS) and wall model for large-eddy simulation
(LES) is proposed by devising the flow as a collection of building
blocks that enables the prediction of the eddy viscosity. The core
assumption of the model is that simple canonical flows contain the
essential physics to provide accurate predictions of the SGS tensor in more complex
flows. The model is constructed to predict zero-pressure-gradient
wall-bounded turbulence, adverse pressure gradient effects,
separation and laminar flow. The approach is implemented using a Bayesian classifier, which identifies
the contribution of each building block in the flow, and a neural-network-based predictor,
which estimates the eddy viscosity based on the building-block
units. The training data are directly obtained from wall-modeled LES
with an exact SGS/wall model for the mean quantities to guarantee
consistency with the numerical discretization. The model is validated
in canonical flows and the NASA High-Lift Common Research Model and shown to improve the
predictions with respect to current modeling approaches.

\section{Introduction}
  
The ability to accurately simulate complex flows using computational
fluid dynamics (CFD) is crucial in many applications ranging from
aircraft design to drag reduction in
pipelines~\citep{smits2013wall}. In recent years, wall-modeled
large-eddy simulation (WMLES) has gained traction as an
alternative to traditional lower-fidelity models, such as
Reynolds-averaged Navier-Stokes (RANS)-based methods. Compared to
RANS-based methods, WMLES is inherently unsteady and better positioned
to provide high accuracy in the presence of non-equilibrium flows such as
separation. Additionally, since WMLES only resolves the large-scale
motions far from the wall, its computational cost is competitive
compared to other CFD
approaches~\citep{choi2012grid,Yang2021}. Recently,
\cite{goc2021large} have shown that WMLES is close to achieving the
accuracy and turnaround times demanded by the aerospace
industry. While state-of-the-art WMLES performs satisfactorily in
turbulent boundary layers with sufficient grid resolution, its
performance deteriorates in the presence of less than 20 grid points
per boundary layer thickness. Unfortunately, the latter grid
resolution is typical for external aerodynamics
applications~\citep{Lozano2022}. In this report,
we present a unified SGS/wall model consistent with the numerical
discretization and applicable to complex geometries.  The model is
rooted in the idea that truly revolutionary improvements in WMLES will
encompass advancements in numerics, grid generation and wall/SGS
modeling.

The model is implemented using artificial neural networks (ANNs) within 
the supervised learning paradigm, which has been extensively explored
for modeling in recent years.  For example, SGS models have been
trained using data from filtered direct numerical simulation
(DNS)~\citep{gamahara2017searching, xie2019artificial}. Examples of
supervised learning for wall modeling include \cite{yang2019predictive}, \cite{zhou2021wall}, \cite{zangeneh2021data} and \cite{huang2019wall}, to name a
few. Interestingly, most studies to date have trained the model with data from higher-fidelity simulations, such as DNS or wall-resolved LES. As a consequence, previous models ignored the
nonnegligible errors arising from the numerical discretization in
actual WMLES. An exception is the work by \cite{bae2022} using
reinforcement learning. Here, we take advantage of a novel data
preparation process to ensure numerical consistency between the
training data and the model deployed in the flow solver.


The report is organized as follows. Section \ref{Methodology} details
the simulation setup, numerical approach, model formulation, control
scheme and training procedure. Results for different test cases are
presented in Section \ref{Results}. These include laminar channel
flow, turbulent channel flow, turbulent Poiseuille-Couette flow and
the NASA High-Lift Common Research Model (CRM-HL). Finally, concluding
remarks are offered in Section \ref{Conclusions}.

\section{Methodology}
\label{Methodology}

\subsection{Model overview}
\label{sub:Model Overview}

The model presented is based on the hypothesis that complex flows can
be divided into subregions resembling canonical building-block
flows. Following the strategy proposed by \cite{lozanoself}, we devise
a model composed of a classifier and a predictor. The classifier
categorizes the flow into different types of canonical flows, and the
predictor provides the wall stress and eddy viscosity based on
the likelihood of each category. The model is referred to as the BFM
(building-block flow model). In this work, we consider four types of
building-block flows, viz., laminar channel flow, fully developed turbulent channel
flow, turbulent Poiseuille-Couette flow with adverse pressure
gradient and turbulent Poiseuille-Couette flow with ``separation''.

\begin{figure}
  \centering
        \ig[width=\tw,trim=0 3cm 0 2cm, clip]{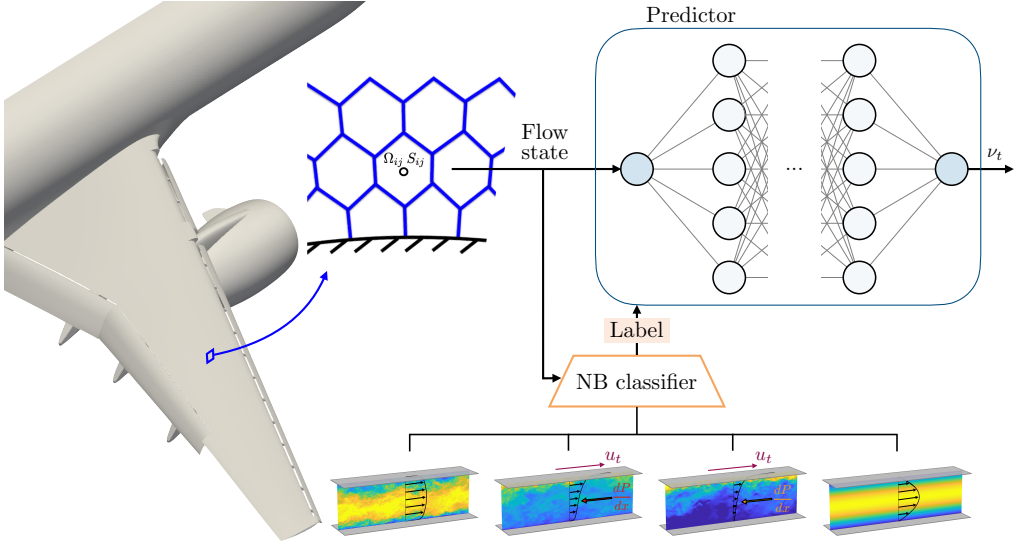}
    \caption{Schematic of the building-block flow model. The panel shows the classifier-predictor structure and the ANN architecture. The flow state refers to $(I_1,...,I_6,\nu,\Delta,u_{||})$ and different ANNs are used for the inner and outer layer. The bottom of the figure depicts the building-block flows considered (from left to right): turbulent channel flow, Poiseuille-Couette flow with mild adverse pressure gradient, Poiseuille-Couette flow with separation and laminar Poiseuille flow.}
    \label{fig:Model}
\end{figure}

The model architecture is summarized in Figure~\ref{fig:Model}. The
model inputs are the invariants of rate-of-strain and rate-of-rotation
tensors ($\bar{S}_{ij}$ and $\bar{R}_{ij}$, respectively), which are
defined as
\begin{equation}
  \begin{array}{ll}
    I_1=\operatorname{tr}\left(\mathbf{\bar{S}}^2\right), & I_2=\operatorname{tr}\left(\mathbf{\bar{R}}^2\right), \\
    I_3=\operatorname{tr}\left(\mathbf{\bar{S}}^3\right), & I_4=\operatorname{tr}\left(\mathbf{\bar{S} \bar{R}}{ }^2\right), \\
    I_5=\operatorname{tr}\left(\mathbf{\bar{S}}^2 \mathbf{\bar{R}}^2\right), & I_6=\operatorname{tr}\left(\mathbf{\bar{S}}^2 \mathbf{\bar{R}}^2 \mathbf{\bar{S} \bar{R}}\right),
  \end{array}
\end{equation}
where $\bar{S}_{ij}$ and $\bar{R}_{ij}$ are
\begin{equation}
  \begin{aligned}
\bar{S}_{i j}=\frac{1}{2}\left(\frac{\partial \bar{u}_i}{\partial x_j}+\frac{\partial \bar{u}_j}{\partial x_i}\right) , \
\bar{R}_{i j}=\frac{1}{2}\left(\frac{\partial \bar{u}_i}{\partial
x_j}-\frac{\partial \bar{u}_j}{\partial x_i}\right).
  \end{aligned}
\end{equation}
%
The anisotropic component of the resolved shear stress is given by the
eddy-viscosity model
\begin{equation}
  \bar{\tau}_{ij}^d = -2\nu_t \bar{S}_{ij},
  \label{eq:eddy}
\end{equation}
and the eddy viscosity is assumed to be a function of the
invariants~\citep{lund1993parameterization}
\begin{equation}
  \nu_t = f(I_1,I_2,I_3,I_4,I_5,I_6, \theta),
  \label{eq:map}
\end{equation}
where $f$ represents an ANN, and $\theta$ denotes additional input
variables, namely, $\nu$, $\Delta$ and $u_{\parallel}$, where $\nu$
is the kinematic viscosity, $\Delta=\sqrt{\Delta x^2+\Delta y^2+\Delta
  z^2}$ is the characteristic grid size and $u_{\parallel}$ is the
magnitude of the wall-parallel velocity measured with respect to the wall. The latter is only used for
the ANN close to the wall. The mapping between inputs and output is
learned from data generated from controlled WMLES simulations, which
are detailed in Section~\ref{sub:Building-blocks}. 
%
%

The input and output variables of the ANN are given in nondimensional
form.  This is crucial for the generalizability of the model. 
The nondimensionalization of the input and output features is done
using parameters that are local in both time and space
to guarantee the
applicability of the model to complex geometries. Two types of ANNs
are considered. For grid points at the wall
and the first grid points off the wall, the input and output
quantities are nondimensionalized using viscous scaling ($\nu$ and
$\Delta$). For the rest of the points, the input and output variables
are nondimensionalized using semi-viscous scaling
[$(\sqrt{2\bar{S}_{ij}\bar{S}_{ij}} \nu)^{1 / 2}$
and $\Delta$].

The simulations for canonical cases are performed using the
solver by \cite{bae2021effect} (i.e., the incompressible, staggered,
second-order accurate central finite difference method). The CRM-HL
simulation is carried out in charLES, a finite-volume, compressible
flow solver. Readers are referred to \cite{goc2021large} for more
details about charLES.


\subsection{Building blocks and data preparation}
\label{sub:Building-blocks}

Figure~\ref{fig:Model} also illustrates the collection of building-block
flows.  The Poiseuille flow is used to generally represent laminar
flows. The turbulent channel flow models the regime where turbulence
is fully developed without significant mean-pressure gradient effects. In both
cases, the bottom and top walls are static.  In the turbulent
Poiseuille-Couette flows, the top wall moves at a constant speed ($u_t$) in
the streamwise direction, and an adverse pressure gradient is applied
in the direction opposed to the top wall velocity. The adverse
pressure gradient ranges from mild to strong, so that the flow
separates (i.e., zero wall stress) on the bottom wall.
%

The training data is generated using WMLES with an exact-for-the-mean
eddy-viscosity and wall model (denoted by ESGS). This enables the
model to account for the numerical errors of the flow solver. The ESGS
model is based on the anisotropic minimium-dissipation (AMD) SGS
model~\citep{rozema2015minimum} combined with a controller that
iteratively adjusts the AMD eddy viscosity to match the DNS mean
velocity profile. A Dirichlet non-slip boundary condition is applied
at the walls and the correct wall-shear stress is enforced by
augmenting the eddy viscosity at the walls such that
$\left.\nu_t\right|_w=\left.\left(\partial \bar{u}/\partial
y\right)\right|_w ^{-1} \tau_w/\rho-\nu$,
%
%
following \cite{bae2021effect}.  For turbulent channel flows, the mean
DNS quantities are obtained from the database by Jimen\'ez and
coworkers \citep{del2004scaling, hoyas2006scaling, lozano2014effect}.  Four cases with friction
Reynolds numbers 550, 950, 2000 and 4200 are used.  For the turbulent
Poiseuille-Couette flows, our in-house code is used to generate DNS
data. Two cases (labeled as PC-0 and PC-2) are considered, which
correspond to separation and adverse
pressure gradient, respectively. The Reynolds numbers based on the
adverse pressure gradient are $\mathit{Re_{P}} =
{\sqrt{h^3\mathrm{d}P/\mathrm{d}x}}/\nu=$ 680 and 340, and the Reynolds
number based on the top wall velocity is $\mathit{Re_U} = u_t h
/\nu=22,360$. The computational domain is $L_x\times L_y \times
L_z=4\pi h\times2h\times 2\pi h$ for channel flows and $L_x\times L_y
\times L_z=2\pi h\times2h\times \pi h$ for PC-0 and PC-2, where $h$ is
the channel half-height. The grid size for the WMLES cases with ESGS
is $\Delta \approx 0.2h$.

%

\subsection{Predictor and classifier}
\label{sub:ANN}

The architecture of the ANNs is a fully connected feedforward neural
network with 6 hidden layers and 40 neurons per layer. The model comprises different ANNs. The
first ANN is used to predict the eddy viscosity far from the wall,
where viscous effects are negligible (i.e., outer layer ANN).  To
account for near-wall viscous effects, two ANNs are used to predict
$\nu_t$ at the wall and at the first grid point off the wall.  
The input features to the outer layer ANN are 
$(I_1,...,I_6,\nu,\Delta)$. For inner layer ANNs, the input features are
identical to the outer layer ANN in addition to $u_{||}$. The
ANNs were trained using stochastic gradient descent by randomly
dividing training data into two groups, the training set (80\% of the
data) and test set (20\% of the data). For laminar flows, the eddy
viscosity is set to zero.

The classification of the flow at a given point is done in two
steps. First, the flow is classified as turbulent
or laminar. If the
flow is turbulent, it is further classified as zero-pressure-gradient
wall turbulence (ZPG), adverse-pressure-gradient wall turbulence (APG), or separated turbulent flow.  A Naive Bayes (NB) classifier is
employed whose inputs are the nondimensionalized invariants (together with $u_\parallel$ normalized with $\nu / \Delta$ for the second classification).  For
the turbulent/laminar classifier, the invariants are normalized with
$\nu$ and $\Delta$, whereas for the second classification, $\sqrt{2\bar{S}_{ij}\bar{S}_{ij}}$, is used.
The output of the classifier is a label that is fed to the predictor, 
as shown in Figure~\ref{fig:Model}.

\begin{figure}
  \begin{center}
      \ig[width=\tw]{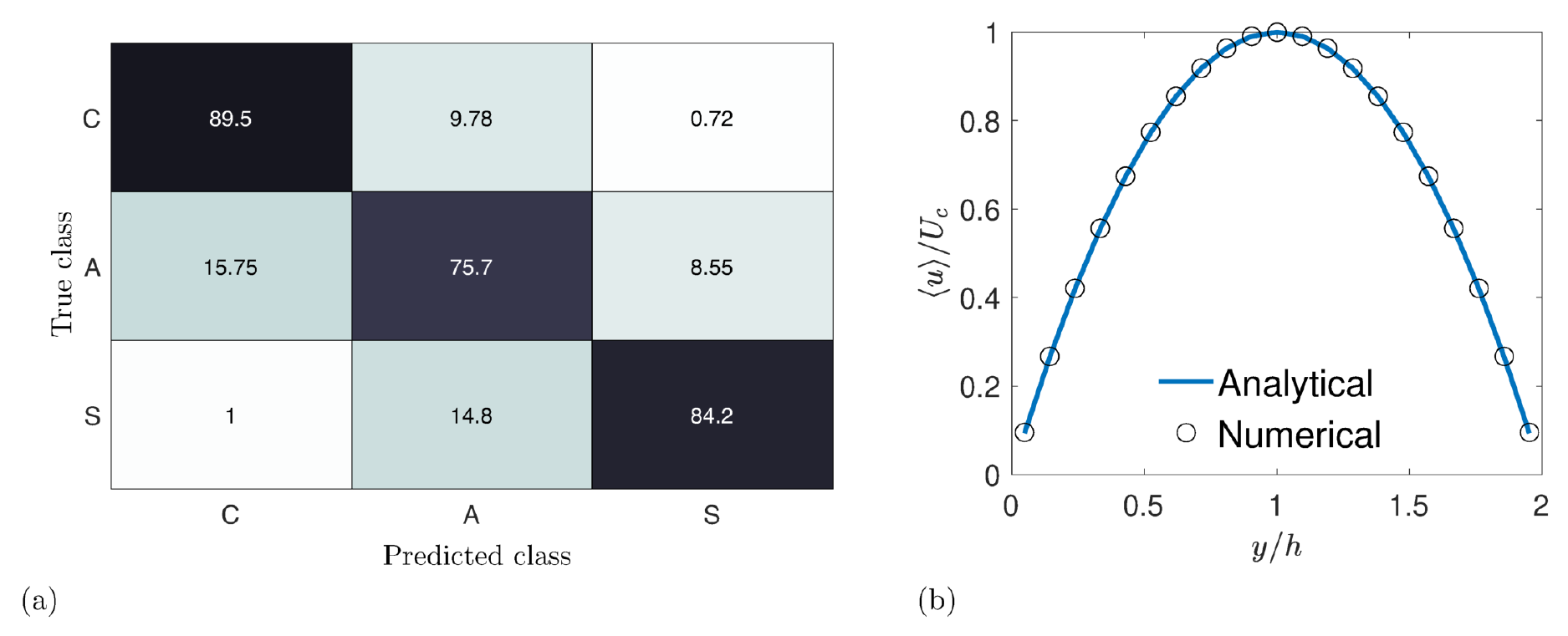}
      \caption{(a) Confusion matrix of the classifier for the first
        grid point off wall. C stands for ZPG flow, A for APG flow and S for separated flow. (b) Mean streamwise
        velocity profile for laminar channel flow ($U_c$ is the
        centerline velocity).\label{fig:class_lam}}
  \end{center}
\end{figure}

The classifier is trained with the same cases as the predictor in
addition to synthetic data generated for laminar Poiseuille flow at
$\mathit{Re_\tau} = [5 - 150]$.  The classifier is applied to all grid points
to discern between turbulent and laminar flow.  The second
classification is only applied to the points in contact with the
wall. The classifier shows 100\% accuracy in the first classification.
For the turbulent classification, Figure~\ref{fig:class_lam}(a) shows
the confusion matrix.  The numbers in the cells indicate the
percentage of samples that are classified as a given class, 
showing that flows are predicted with $>75\%$ accuracy.
%

\section{Results}
\label{Results}
\subsection{Laminar channel flow} 
\label{sub:Laminar Channel Flow}

The BFM is first tested in a laminar channel flow with fixed
centerline velocity.  The simulation is started from a random flow
field and the data are collected after transients.  As shown in
Figure~\ref{fig:class_lam}(b), the BFM produces the same solution as the analytical one by correctly classifying all the points as laminar flow. Regarding wall-shear stress, the analytical solution corresponds to a friction Reynolds number $\mathit{Re_{\tau}}=19.64$, and the prediction of BFM
is $\mathit{Re_{\tau}}=19.62$. 

\subsection{Turbulent channel flow} 

\begin{figure}
  \begin{center}
    \includegraphics[width=.75\tw]{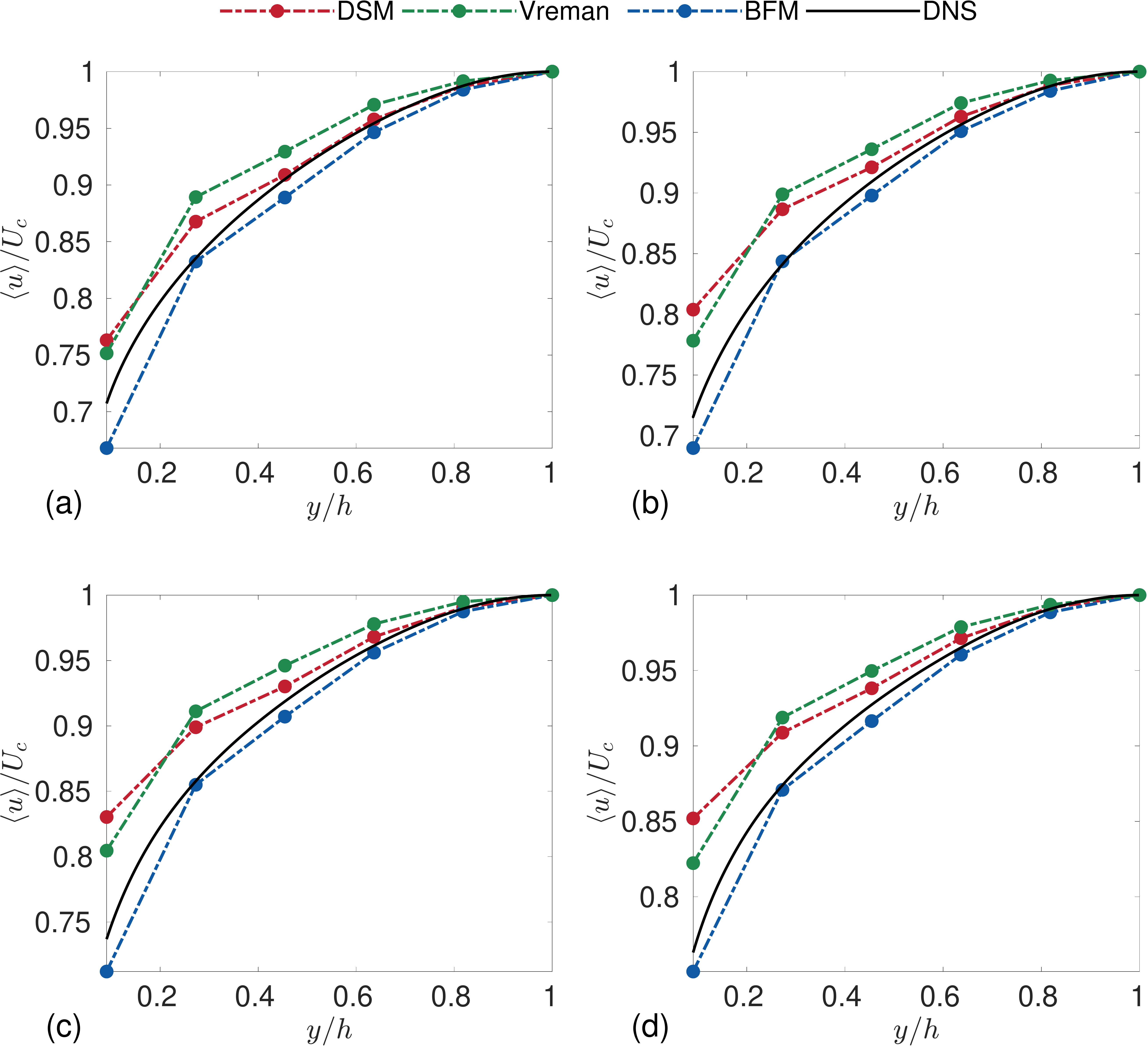}
  \end{center}
    \caption{Mean streamwise velocity for turbulent channel flow using
      a coarse grid ($\Delta=0.2h$). (a) $\mathit{Re_{\tau}}\approx550$, (b) $\mathit{Re_{\tau}}\approx950$, (c) $\mathit{Re_{\tau}}\approx2000$ and (d) $\mathit{Re_{\tau}}\approx4200$.}\label{fig:turbCH}
\end{figure}

\begin{figure}

  \begin{center}
    \includegraphics[width=.75\textwidth]{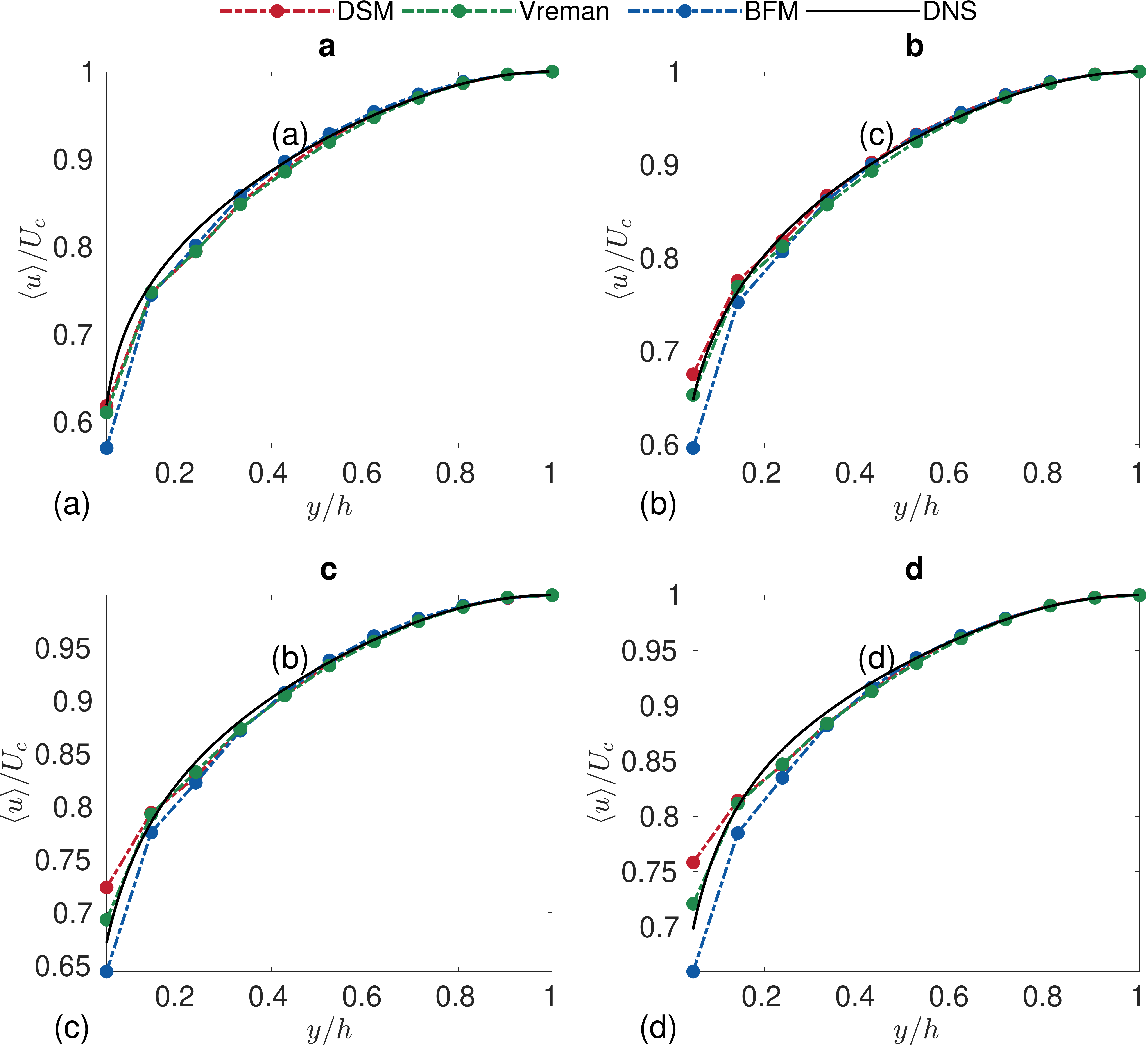}
  \end{center}
    \caption{Mean streamwise velocity for turbulent channel flow using
      a medium grid ($\Delta=0.1h$). (a) $\mathit{Re_{\tau}}\approx550$, (b) $\mathit{Re_{\tau}}\approx950$, (c) $\mathit{Re_{\tau}}\approx2000$ and (d) $\mathit{Re_{\tau}}\approx4200$.}
    \label{fig:turbCHN}
\end{figure}

The BFM is next tested in turbulent channel flows. Two grid sizes
($\Delta = 0.2h$ and $\Delta = 0.1h$) are considered to assess the
generalizability of the model. The results of BFM are compared with
the Dynamic Smargorinsky (DSM) and Vreman SGS models using the
ODE-based equilibrium wall model (EQWM) by \cite{larsson2016large}.
The mean velocity profiles are shown in Figure~\ref{fig:turbCH}, and
Table~\ref{tab:Re} summarizes the friction Reynolds number predicted
by WMLES. BFM exhibits the best performance in the coarse grid. The
predictions of BFM deteriorate when the grid size is refined as shown
in Figure~\ref{fig:turbCHN}; however, they are still comparable to those from
traditional models. In this preliminary version of BFM, the model was trained only at one grid resolution.  Better
grid-convergence behavior is expected in future versions of BFM
trained at various grid sizes.
\label{sub:Turbulent Channel Flow}
\begin{table}
\begin{center}
\resizebox{.85\textwidth}{!}{\begin{tabular}{lllllcc}
    Case & $Re_{\tau}$ & DSM-EQWM &  Vreman-EQWM & BFM & Trained & $Re_P$\\
    \hline
    550-C & 547 & 587(7.3\%)  & 578(5.7\%) & \look{540(1.3\%)}    & \checkmark & N/A\\
    950-C & 943 & 1034(9.7\%)  & 1002(6.3\%) & \look{929(1.5\%)}     & \checkmark & N/A\\
    2000-C & 2003 & 2228(11.2\%)  & 2157(7.7\%) & \look{1976(1.3\%)}  & \checkmark & N/A\\
    4200-C & 4179 & 4641(11.1\%)  & 4474(7.1\%) & \look{4179(0.0\%)}  & \checkmark & N/A\\
    550-M & 547 & 552(0.9\%)  & \look{545(0.4\%)} & 537(1.8\%)    & \tikzxmark & N/A\\
    950-M & 943 & 973(3.2\%)  & \look{944(0.1\%)} & 906(3.9\%)     & \tikzxmark & N/A\\
    2000-M & 2003 & 2137(6.7\%)  & 2050(2.3\%) & \look{1988(0.7\%)}  & \tikzxmark & N/A\\
    4200-M & 4179 & 4502(7.7\%)  & \look{4286(2.6\%)} & 4006(4.1\%)  & \tikzxmark & N/A\\
    PC0 & 7 & 76(1014\%)  & 68(871\%) & \look{33(371\%)}  & \checkmark & 680\\
    PC1 & 194 & 164(15.5\%)  & 169(12.9\%) & \look{172(11.3\%)}  & \tikzxmark & 480\\
    PC2 & 264 & 225(14.8\%)  & 221(16.3\%) & \look{227(14.0\%)}  & \checkmark & 340\\
    \end{tabular}}
\end{center}
\caption{Friction Reynolds numbers predicted by different SGS and wall
  models.  $\mathit{Re_{\tau}}$ are the benchmark values from DNS. The ``Trained'' column
  shows whether the case has been used for training. Cases ended with M were simulated with a finer grid ($\Delta=0.1h$). Numbers in
  parentheses are the relative errors defined by $\left|(\mathit{Re_{\tau,
    \text{pred}}} - \mathit{Re_{\tau}})/\mathit{Re_{\tau}} \right|$. $\mathit{Re_P}$ is the
  Reynolds number defined by the adverse pressure gradient applied.}
\label{tab:Re}
\end{table}

\subsection{Turbulent Poiseuille-Couette flow} 
\label{sub:Poiseuille-Couette Flow}

\begin{figure}
  \begin{center}
    \includegraphics[width=.9\textwidth]{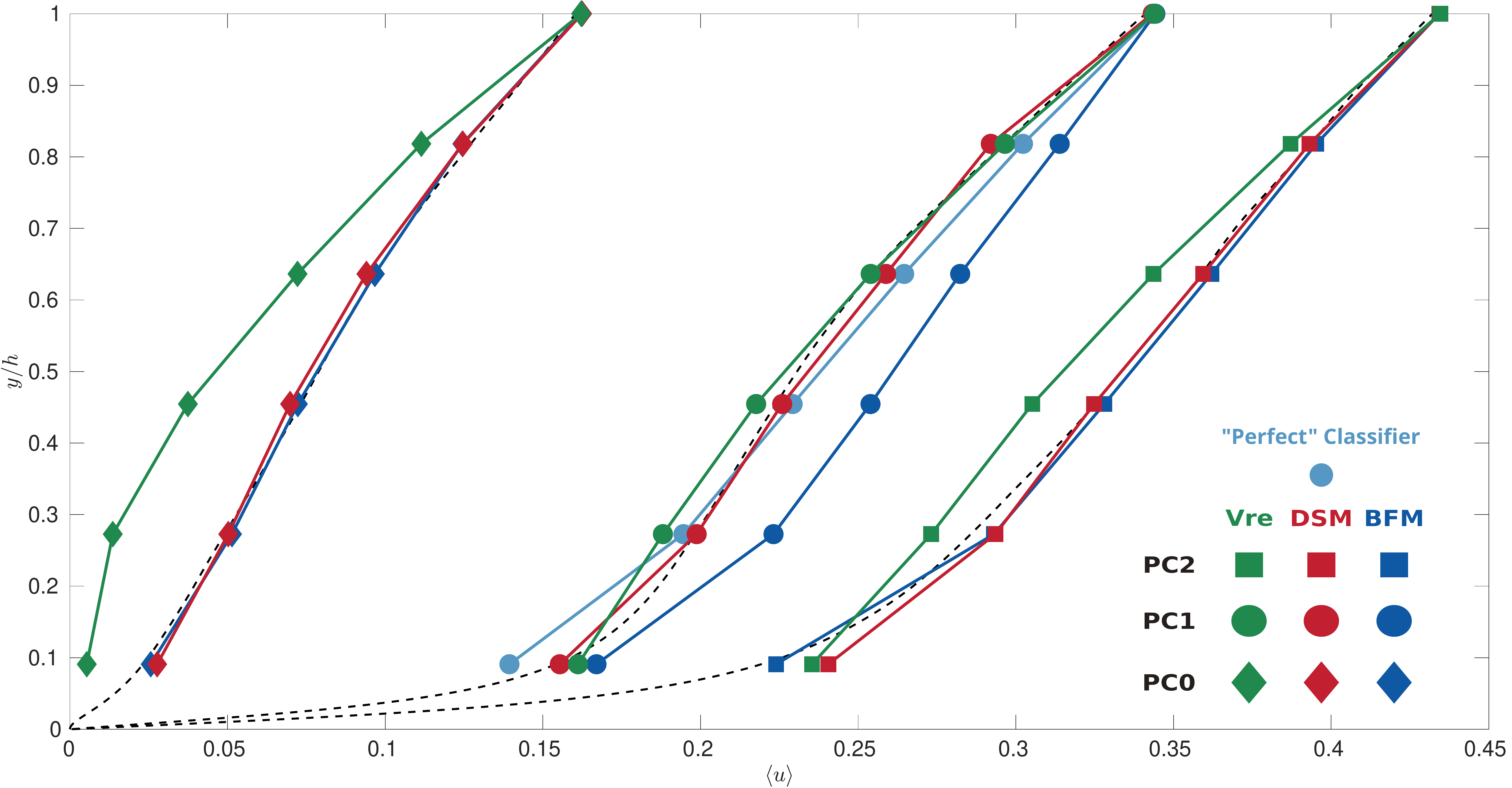}
    \caption{Mean streamwise velocity for Poiseuille-Couette
      flow. Three cases, PC0 (separation), PC1 (stronger APG) and PC2
      (mild APG), are overlaid on the same plots.}
    \label{fig:PC}
  \end{center}
\end{figure}

Three cases are considered: PC0 (separation), PC1 (stronger APG) and
PC2 (mild APG).  The mean velocity profiles and wall-shear stress
predictions are shown in Figure~\ref{fig:PC} and Table~\ref{tab:Re},
respectively. As shown in
Table~\ref{tab:Re}, BFM gives the best wall-stress predictions for all cases.  For PC0, BFM is capable of
capturing separation better than other models.  For PC2 and PC1, the improvement
is marginal. As shown in Figure~\ref{fig:PC}, BFM also outperforms or matches 
the accuracy of other models in terms of the prediction of the mean velocity 
profile except for PC1. The classifier was found to be responsible for hindering the overall performance in this case. Assuming a  perfect classification for PC1, the results by BFM outperform other models, as shown in Figure~\ref{fig:PC}. A more accurate classifier is
anticipated to help the model achieve better mean profile and wall-shear stress predictions.

\subsection{CRM-HL} 
\label{sub:CRM}

Finally, BFM is tested in the CRM-HL as representative of
realistic aircraft in high-lift configuration~\citep{lacy2016development}. We
focus on free air simulations. The reader
is referred to \cite{goc2022large} for a detailed description
 of the physical and numerical setup.

\begin{figure}
      \begin{center}
        \ig[width=.9\textwidth]{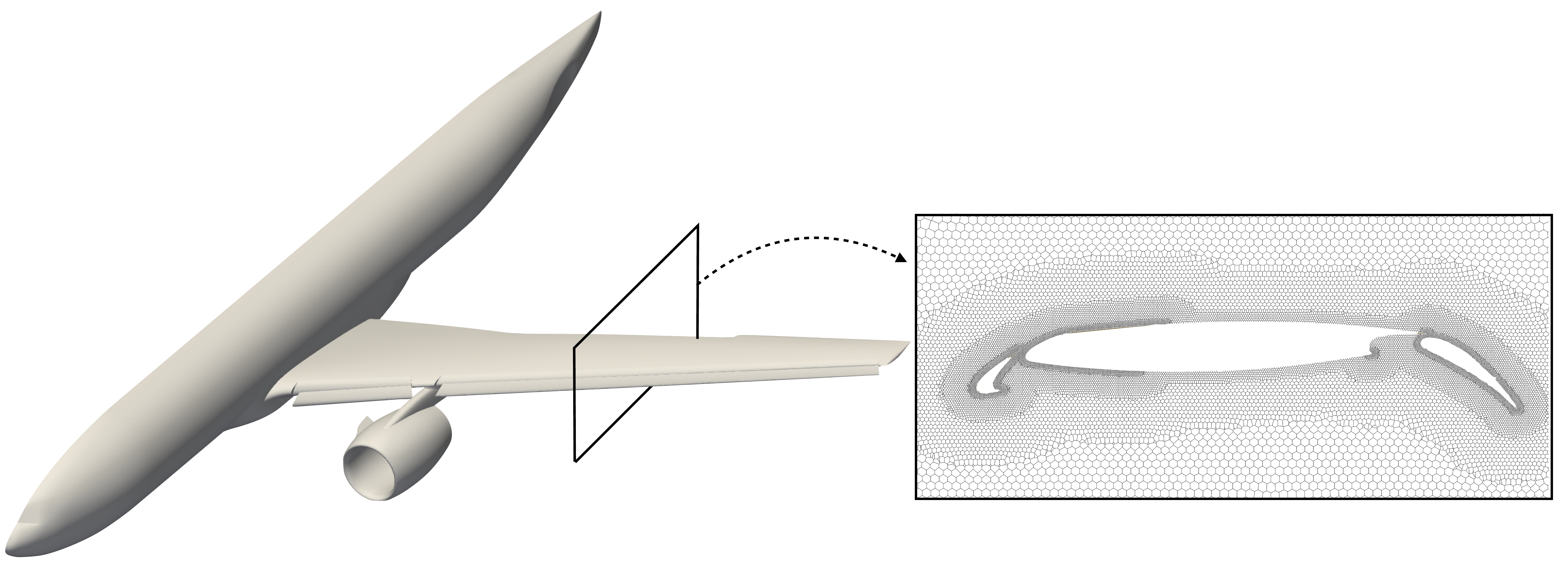}
        \caption{Geometry of the CRM-HL model and a cross-sectional
          view of the grid over the wing surface.\label{fig:CRMG}}
  \end{center}
\end{figure}

We perform simulations for two angles of attack, $\alpha = 7.05^\circ$ and
$19.57^\circ$, and compare the lift, drag and pitching moment
coefficients with experimental data and simulations from
\cite{goc2022large} using DSM with EQWM (labeled as DSM-EQWM).  The
grid has 40 million elements and is the same as in
\cite{goc2022large}.  A slice of the grid is depicted in
Figure~\ref{fig:CRMG}. The simulations are conducted in charLES with a
version of BFM that assumes ZPG. Future versions of
BFM will be retrained using WMLES data from charLES to guarantee
accuracy in the classification and consistency with the numerical
discretization.

\begin{figure}
    \begin{center}
        \ig[width=\tw]{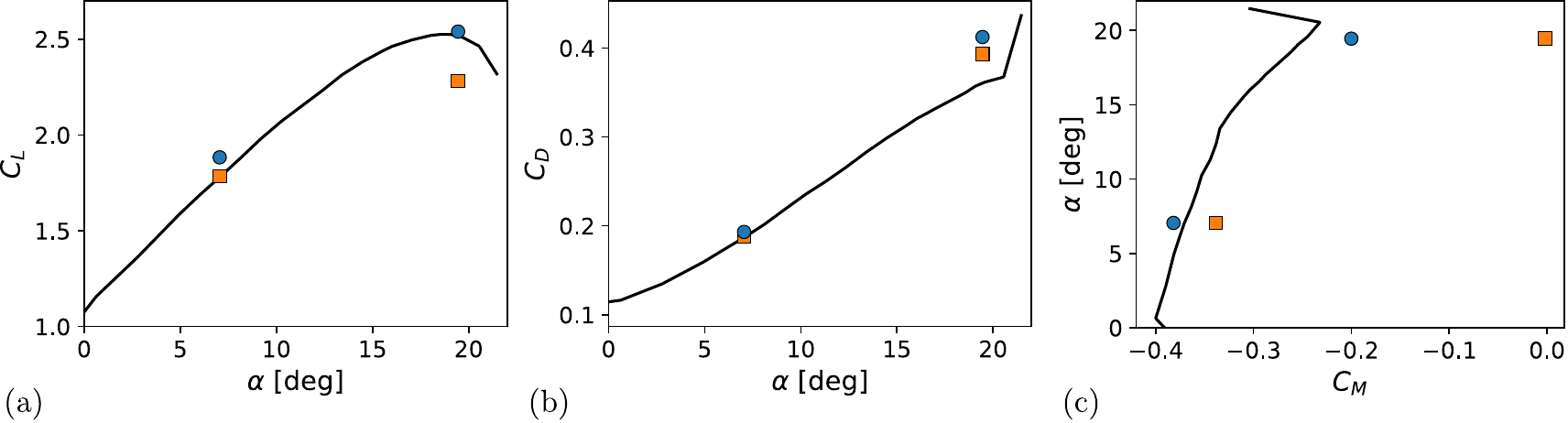}
    \caption{(a) Lift, (b) drag and (c) pitching moment coefficients
      for CRM-HL. The black lines denote experimental results, squares are for DSM-EQWM and circles are for BFM.}
    \label{fig:CRMForces}
  \end{center}
\end{figure}

\begin{figure}
    \begin{center}
        \ig[width=.8\tw]{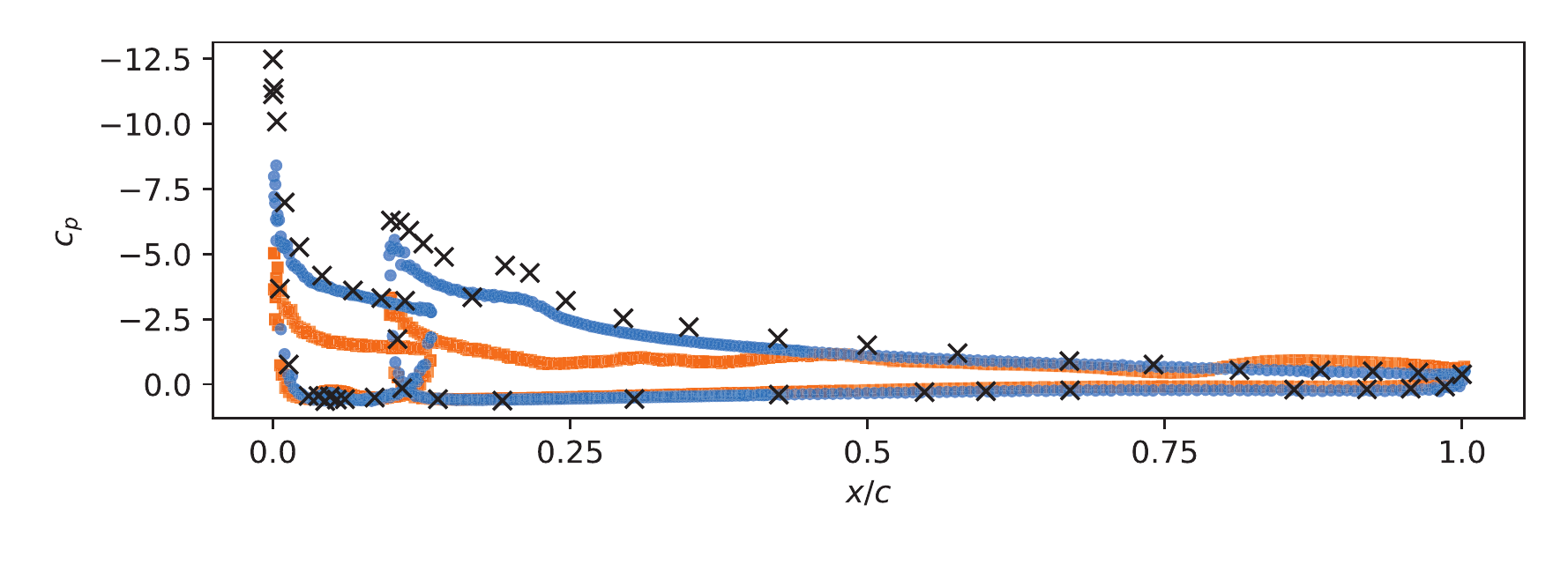}
        \caption{Pressure coefficient at the 82\% spanwise section 
        of the wing for $\alpha = 19.57^\circ$. Black $\times$ represents experimental results, orange squares are for DSM-EQWM and blue circles are for BFM.\label{fig:CRMcp}}
  \end{center}
\end{figure}

Figure~\ref{fig:CRMForces} displays the lift, drag and pitching
moment coefficients.  For $\alpha = 7.05^{\circ}$, DSM-EQWM correctly predicts
$C_L$ and $C_D$. However, further analysis has shown that this
accurate prediction is coincidental and due to error cancellation when
integrating the total forces. This is corroborated by the poor
predictions of pitching moment by DSM-EQWM. However, BFM
provides more consistent results.  The case of $\alpha = 19.57^{\circ}$
corresponds to the angle of attack of maximum lift
coefficient. DSM-EQWM underpredicts the $C_L$, whereas BFM accurately
predicts $C_L$.  The drag coefficient is overpredicted by both models,
but the pitching moment computed using BFM is closer to the
experimental values, suggesting that BFM provides a more accurate
distribution of the forces over the wing. Inspection of the sectional
pressure coefficient (Figure~\ref{fig:CRMcp}) indicates that BFM maintains the flow
attached over a longer section of the wing compared to DSM-EQWM, which
explains the enhanced performance of BFM.
\section{Conclusions}
\label{Conclusions}

We have presented a unified SGS/wall model for large-eddy simulation by
devising the flow as a collection of building blocks whose information
enables the prediction of the SGS stress tensor. The model is referred
to as the building-block flow model (BFM) and is grounded in the idea that
truly revolutionary improvements in WMLES will encompass advancements
in machine-learning-based modeling consistent with the numerics and
grid generation. 

The first version of BFM discussed here is constructed 
to predict wall-attached turbulence, adverse-pressure-gradient turbulence,
separation and laminar flow. The model is trained using WMLES with an
exact-for-the-mean SGS/wall model that guarantees consistency with 
the numerical schemes and the gridding strategy. The model is applicable 
to complex geometries and is implemented using artificial neural 
networks. 

The BFM has been validated in canonical flows. These include laminar Poiseuille flow, turbulent channel flows, and turbulent Poiseuille-Couette flows mimicking adverse-pressure-gradient effects and separation.  The performance of BFM in complex scenarios 
is evaluated in the NASA Common Research Model High-Lift (CRM-HL). We have shown that 
BFM outperforms traditional SGS/wall models in the canonical flows and the CRM-HL. The only exceptions are the coincidentally accurate predictions by the 
equilibrium wall model due to error cancellation. To the best of our knowledge, 
this is the first demonstration of a successful SGS/wall model  
implemented via artificial neural networks applicable to realistic aircraft 
configurations.

\subsection*{Acknowledgments}

The authors acknowledge the MIT
SuperCloud and Lincoln Laboratory Supercomputing Center for providing
HPC resources that have contributed to the research results reported
within this paper.

\bibliographystyle{ctr}

\end{document}